\documentclass[useAMS,usenatbib]{mn2e}
\usepackage{epsfig}
\usepackage{amsmath}

\newcommand{\be}{\begin{equation}}
\newcommand{\beq}{\begin{equation}}
\newcommand{\ba}{\begin{eqnarray}}
\newcommand{\ee}{\end{equation}}
\newcommand{\eeq}{\end{equation}}
\newcommand{\ea}{\end{eqnarray}}

\newcommand{\apj}{ApJ}
\newcommand{\apjl}{ApJL}
\newcommand{\mnras}{MNRAS}
\newcommand{\aj}{AJ}
\newcommand{\apjs}{ApJS}

\def\lsim{~\rlap{$<$}{\lower 1.0ex\hbox{$\sim$}}}

\def\gsim{~\rlap{$>$}{\lower 1.0ex\hbox{$\sim$}}}

\title[21cm emission after reionization]{Fluctuations in 21cm Emission After Reionization}

\author[Wyithe \& Loeb]{J. Stuart B. Wyithe$^1$ and Abraham Loeb$^2$\\$^1$
School of Physics, University of Melbourne, Parkville, Victoria,
Australia\\$^2$ Harvard-Smithsonian Center for Astrophysics, 60 Garden St.,
Cambridge, MA 02138\\Email: swyithe@physics.unimelb.edu.au,
loeb@cfa.harvard.edu}

\begin{document}

%\date{\today}
%\pagerange{\pageref{firstpage}--\pageref{lastpage}} \pubyear{2006}

\maketitle

\label{firstpage}
\begin{abstract}

The fluctuations in the emission of redshifted 21cm photons from neutral
inter-galactic hydrogen will provide an unprecedented probe of the
reionization era.  Conventional wisdom assumes that this 21cm signal
disappears as soon as reionization is complete, when little atomic hydrogen
is left through most of the volume of the inter-galactic medium
(IGM). However observations of damped Ly$\alpha$ absorbers indicate that the
fraction of hydrogen in its neutral form is significant by mass at all
redshifts.  Here we use a physically-motivated model to show that residual
neutral gas, confined to dense regions in the IGM with a high recombination
rate, will generate a significant post-reionization 21cm signal. We show
that the power-spectrum of fluctuations in this signal will be detectable
by the first generation of low-frequency observatories at a signal-to-noise
that is comparable to that achievable in observations of the reionization era.  The
statistics of 21cm fluctuations will therefore probe not only the
pre-reionization IGM, but rather the entire process of HII region overlap,
as well as the appearance of the diffuse ionized IGM.

\end{abstract}

\begin{keywords}
cosmology: diffuse radiation, large scale structure, theory -- galaxies: high redshift, inter-galactic medium
\end{keywords}

\section{Introduction}

The reionization of cosmic hydrogen by the first stars and quasars (e.g. Barkana
\& Loeb~2001), was an important milestone in the history of the Universe.
The recent discovery of distant quasars has allowed detailed absorption
studies of the state of the high redshift intergalactic medium (IGM) at a
time when the universe was less than a billion years old (Fan et al.~2006;
White et al.~2003).  Several studies have used the evolution of the
ionizing background inferred from these spectra to argue that the
reionization of cosmic hydrogen was completed just beyond $z\sim6$ (Fan et
al.~2006; Gnedin \& Fan~2006; White et al.~2003). However, other authors
have claimed that the evidence for this rapid change becomes significantly
weaker for a different choice of density distribution in the IGM (Becker et
al.~2007). Different arguments in favour of a rapidly evolving IGM at $z>6$
are based on the properties of the putative HII regions inferred around
the highest redshift quasars (Wyithe \& Loeb~2004; Wyithe, Loeb, \& Carilli
2005; Mesinger \& Haiman 2005). However, Bolton \& Haehnelt (2007a) and Lidz et
al. (2007) have demonstrated that the interpretation of the
spectral features is uncertain and that the observed spectra could either
be produced by an HII region, or by a classical proximity zone.  One
reason for the ambiguity in interpreting these absorption spectra is that
Ly$\alpha$ absorption can only be used to probe neutral fractions that are
smaller than $10^{-3}$ owing to the large cross-section of the Ly$\alpha$ resonance.

A better probe of the process of reionization is provided by redshifted
21cm observations.  Reionization starts with ionized (HII) regions around
galaxies, which later grow to surround groups of galaxies. The process of
reionization is completed when these HII regions overlap (defining the
so-called {\it overlap} epoch) and fill-up most of the volume between
galaxies.  The conventional wisdom is that the 21cm signal disappears after
the {\it overlap} epoch, because there is little neutral hydrogen left
through most of intergalactic space. However, the following simple estimate
can be used to demonstrate that 21cm emission should be significant even
after reionization is completed. Damped Ly$\alpha$ systems are believed to
contain the majority of the neutral gas at high redshifts. Indeed,
observations of damped Ly$\alpha$ systems out to a redshift of $z\sim4$
show the cosmological density parameter of HI to be $\Omega_{\rm
HI}\sim10^{-3}$ (Prochaska et al ~2005). In the standard cosmological model
the density parameter of baryons is $\Omega_{\rm b}\sim 0.04$, so that the
mass-averaged neutral hydrogen fraction at $z\sim4$ (long after the end of
the HII overlap epoch) is $f_{\rm m}\sim0.03$. This neutral gas does not
contribute significantly to the effective Ly$\alpha$ optical depth, which
is sensitive to the volume averaged neutral fraction (with a value that is
orders of magnitude lower). However the redshifted 21cm emission is
sensitive to the total (mass-weighted) optical depth of this neutral gas.
Observations of the redshifted 21cm signal would therefore detect the total
neutral hydrogen content in a volume of IGM dictated by the observatory
beam and frequency band-pass.

Although the 21cm emission after HII overlap is dominated by dense clumps
of gas rather than by diffuse gas in the IGM as is the case before
reionization is complete, we do not expect 21cm self absorption to impact
the level of 21cm emission. This conclusion is based on 21cm absorption
studies towards damped Ly$\alpha$ systems at a range of redshifts between
$z\sim0$ and $z\sim3.4$, which show optical depths to absorption of the
back-ground quasar flux with values less than a few percent (Kanekar \&
Chengalur 2003; Curran et al. 2007).  The small optical depth for self
absorption is also supported by theoretical calculations of the 21cm
optical depth of neutral gas in high redshift mini-halos (Furlanetto \&
Loeb~2002).  Moreover, damped Ly$\alpha$ systems have a spin temperature
that is large relative to the temperature of the cosmic microwave
background radiation, and will therefore have a level of emission that is
independent of the kinetic gas temperature (e.g. Kanekar \& Chengalur
2003).

At $z\sim4$, the brightness temperature contrast of redshifted 21cm
emission will be $\Delta T\sim0.5$mK. Moreover on the $R\sim10$ co-moving
Mpc (hereafter cMpc) scales relevant for upcoming 21cm experiments, the
{\it root-mean-square} amplitude of density fluctuations at $z\sim4$ is
$\sigma\sim0.2$. Hence, we expect fluctuations in the radiation field of
$\sim0.1$mK on 10~cMpc scales.  The fluctuations in the 21cm emission
signal after reionization are therefore expected to be only an order of
magnitude or so smaller than the largest fluctuations predicted at any time
during the entire reionization era (e.g. Wyithe \& Morales~2007). In addition 
the sky temperature, which provides the limiting factor in the system noise
at the low frequencies relevant for 21cm studies, is proportional to
$(1+z)^{2.6}$, and so is a factor of $\sim 3.4 [(1+z)/5]^{2.6}$ smaller at
low-redshifts than for observations at $z\sim7$.  Thus, detectability of
fluctuations in 21cm emission may not decline substantially following the
overlap epoch, allowing the study of the entire HII overlap process in
redshifted 21cm emission.

In this paper we calculate the mean level of 21cm emission through the
overlap epoch to lower redshifts, as well as the statistics of the
fluctuations about this mean due to patchy reionization and
fluctuations in the density field.  Throughout the paper we adopt the set
of cosmological parameters determined by {\it WMAP3} (Spergel et al. 2007)
for a flat $\Lambda$CDM universe.

\section{Semi-Analytic Model for Reionization}
\label{models}

Miralda-Escude et al.~(2000) presented a model which allows the calculation
of an effective recombination rate in an inhomogeneous universe by assuming
a maximum overdensity ($\Delta_{\rm c}$) penetrated by ionizing photons
within HII regions. The model assumes that reionization progresses rapidly
through islands of lower density prior to the overlap of individual
cosmological ionized regions. Following the overlap epoch, the remaining
regions of high density are gradually ionized. It is therefore hypothesized
that at any time, regions with gas below some critical overdensity
$\Delta_{\rm i}\equiv {\rho_{i}}/{\langle\rho\rangle}$ are highly ionized while
regions of higher density are not. In what follows, we draw primarily from
their prescription and refer the reader to the original paper for a
detailed discussion of its motivations and assumptions.  Wyithe \&
Loeb~(2003) employed this prescription within a semi-analytic model of
reionization. This model was extended by Srbinovsky \& Wyithe~(2006) and by
Wyithe, Bolton \& Haehnelt~(2007).
We refer the reader to those papers and only summarise the basics of
the model below. In the current work we limit our attention to reionization
due to population-II stars (Gnedin \& Fan~2006; Srbinovsky \& Wyithe~2006),
which govern the final stages of reionization even in the presence of an
earlier partial or full reionization by population III stars (e.g. Wyithe
\& Loeb~2003).

Within the model of Miralda-Escude et al.~(2000) we describe the
post-overlap evolution of the IGM by computing the evolution of 
the fraction of mass in regions with overdensity below $\Delta_{\rm i}$,
\begin{equation}
F_{\rm M}(\Delta_{\rm i})=\int_{0}^{\Delta_{\rm i}}d\Delta P_{\rm
V}(\Delta)\Delta,
\end{equation}
where $P_{\rm V}(\Delta)$ is the volume weighted probability distribution for
$\Delta$. In a region of large scale overdensity $\delta$ at $z_{\rm obs}$,
the mass fraction $F_{\rm M}(\Delta_{\rm i})$ [or equivalently $\Delta_{\rm
i}$] therefore evolves according to the equation
\begin{equation}
\label{postoverlap}
\frac{dF_{\rm M}(\Delta_{\rm i})}{dz} =
\frac{1}{n_0}\frac{dn_\gamma(\delta)}{dz}-\alpha_{\rm B}\frac{R(\Delta_{\rm
i})}{a^3}n_{\rm e}\left(1+\delta\frac{D(z)}{D(z_{\rm obs})}\right)\frac{dt}{dz},
\end{equation}
where $D$ is the growth factor of linear density fluctuations, $\alpha_{\rm
B}$ is the case B recombination coefficient, $n_{\rm e}$ is the comoving
electron density, $R(\Delta_{\rm i})$ is the effective clumping factor of
the IGM (see below), and $dn_\gamma/dz$ is the co-moving emissivity of
ionizing photons.  This equation was described in Wyithe \& Loeb~(2003),
but is generalized here and below to regions of large scale overdensity
$\delta$ that differ from the average IGM (Wyithe, Bolton \& Haehnelt~2007).  Integration of
equation~(\ref{postoverlap}) requires knowledge of $P_{\rm
V}(\Delta)$. Miralda-Escude et al.~(2000) quote a fitting function which provides  
 a good fit to the volume weighted probability distribution for the
baryon density in cosmological hydrodynamical simulations. This
probability distribution remains a reasonable description at high
redshift when confronted with a more modern cosmology and updated
simulations, although the addition of an analytical approximation for
the high density tail of the distribution remains necessary as a best
guess at correcting for numerical resolution (Bolton \& Haehnelt~2007b).

Equation~(\ref{postoverlap}) provides a good description of the evolution
of the ionization fraction following the overlap epoch of individual
ionized bubbles, because the ionization fronts are exposed to the mean
ionizing radiation field. However prior to the overlap epoch, the
prescription is inadequate, due to the large fluctuations in the intensity
of the ionizing radiation. A more accurate model to describe the evolution
of the ionized volume prior to the overlap epoch was suggested by
Miralda-Escude et al.~(2000). In our notation the appropriate equation is
\begin{eqnarray}
\label{preoverlap}
\nonumber
\frac{dQ_{\rm i}}{dz} &=& \frac{1}{n^0 F_{\rm M}(\Delta_{\rm
crit})}\frac{dn_{\gamma}(\delta)}{dz}\\
\nonumber
 &-&\left[\alpha_{\rm B}(1+z)^3R(\Delta_{\rm
crit})n_{\rm e}\left(1+\delta\frac{D(z)}{D(z_{\rm obs})}\right)\frac{dt}{dz}\right.\\
 &&\hspace{15mm}+ \left.\frac{dF_{\rm M}(\Delta_{\rm
crit})}{dz}\right]\frac{Q_{\rm i}}{F_{\rm M}(\Delta_{\rm c})}.
\end{eqnarray} 
In this expression, $Q_{\rm i}$ is redefined to be the volume filling
factor within which all matter at densities below $\Delta_{\rm c}$ has been
ionized, and $R$ is the effective clumping of the IGM

Within this formalism, the epoch of overlap is precisely defined as the
time when $Q_{\rm i}$ reaches unity. However, we have only a single
equation to describe the evolution of two independent quantities $Q_{\rm
i}$ and $F_{\rm M}$ (or equivalently $\Delta_{\rm c}$). The relative growth
of these depends on the luminosity function and spatial distribution of the
sources. The appropriate value of $\Delta_{\rm c}$ is set by the mean
separation of the ionizing sources. We assume $\Delta_{\rm c}$ to be
constant with redshift before the overlap epoch and compute results for models with values of $\Delta_{\rm
c}=20$ and $\Delta_{\rm c}=5$. Prior to reionization, these values imply a mean-free path that is comparable to the separation between galaxies (Wyithe \& Loeb~2003). 
Our approach is to compute a reionization history given a
particular value of $\Delta_{\rm c}$, combined with assumed values for the efficiency of star-formation and the fraction of ionizing photons that escape from galaxies. With this history in
place we then compute the evolution of the background radiation field
due to these same sources.  After the overlap epoch, ionizing photons will
experience attenuation due to residual overdense pockets of HI gas.
We use the description of Miralda-Escude et al.~(2000) to estimate the
ionizing photon mean-free-path, and subsequently derive the attenuation of
ionizing photons. We then compute the flux at the Lyman-limit in the IGM due to
sources immediate to each epoch, in addition to redshifted
contributions from earlier epochs.

We next describe our model for the emissivity of population-II
stars. Following Wyithe \& Loeb~(2003) we assume the spectral energy
distribution (SED) of population-II star forming galaxies, using the model
presented in Leitherer et al.~(1999). From this SED we calculate the number
of ionizing photons produced per baryon ($N_\gamma$). Note that we ignore
helium reionization in our modeling. We further assume that ionizing
photons are produced primarily in starbursts, with lifetimes much shorter
than the Hubble time, so that we may express the star formation rate per
unit volume as
\begin{equation}
\frac{d\dot{M}}{dV}(z)=f_\star\frac{dF(\delta,z)}{dt_{\mathrm{year}}}\hspace{2pt}\rho_{\mathrm{b}},
\end{equation}
where $\rho_{\mathrm{b}}$ is the co-moving baryonic mass density, and
$F(\delta,z)$ is the density dependent collapsed fraction of mass in halos
above a critical mass at $z$. The factor $f_\star$ (star formation
efficiency) describes the fraction of collapsed baryonic matter that participates in
star formation. 

In a region of co-moving radius $R$ and mean overdensity $\delta(z)=\delta
[D(z)/D(z_{\rm obs})]$ [specified at redshift $z$ instead of the usual
$z=0$], the relevant collapsed fraction is obtained from the extended
Press-Schechter~(1974) model (Bond et al.~1991) as
\begin{equation}
\label{Fcol}
F(\delta,R,z) = \mbox{erfc}{\left(\frac{\delta_{\rm
c}-\delta(z)}{\sqrt{2\left(\left[\sigma_{\rm
gal}\right]^2-\left[\sigma(R)\right]^2\right)}}\right)},
\end{equation}
where $\mbox{erfc}(x)$ is the error function, $\sigma(R)$ is the variance
of the density field smoothed on a scale $R$, and $\sigma_{\rm gal}$ is the
variance of the density field smoothed on a scale $R_{\rm gal}$.
The latter scale corresponds to the minimum halo mass for starformation, and both variances are evaluated at redshift $z$ rather than at $z=0$.  
In this expression, the critical linear overdensity for the collapse of a
spherical top-hat density perturbation is $\delta_c\approx 1.69$.

\begin{figure*}
\includegraphics[width=15cm]{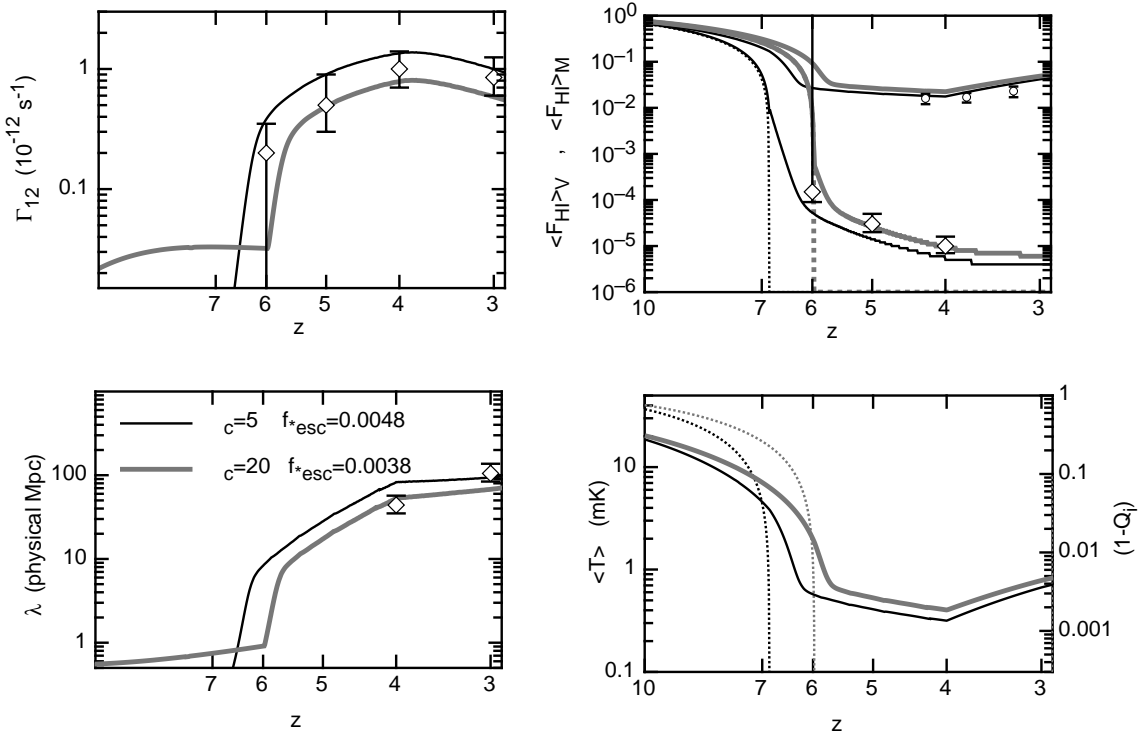} 
\caption{ Models for the reionization of the IGM and the subsequent
post-overlap evolution of the ionizing radiation field. In each panel two cases are
shown, corresponding to different values for the critical overdensity prior
to the overlap epoch ($\Delta_{\rm c}=5$, thin dark lines; and $\Delta_{\rm c}=20$, thick grey lines). We show the cases for the mean IGM with $\delta=0$.
{\em Upper Left Panel:} The ionization rate as a function of redshift. The observational points are from Bolton et
al.~(2007b). {\em Upper Right Panel:} The volume (lower curves) and mass (upper curves)
averaged fractions of neutral gas in the universe. Also shown (dotted
lines) is the fraction of the IGM yet to overlap $(1-Q_i)$. The observational
points for the volume averaged neutral fraction are from Bolton et
al.~(2007b), while the observed mass-fractions are from the damped Ly$\alpha$
measurements of Prochaska et al.~(2005).  {\em Lower Left Panel:} The mean-free-path for ionizing
photons computed using the formalism in \S~\ref{models}. The data points
are based on Storrie-Lombardi et al.~(1994). {\em Lower Right Panel:} The
evolution of the mean 21cm brightness temperature (in mK) with redshift
(solid lines). For comparison, the fraction of IGM yet to overlap $(1-Q_i)$
is overplotted.}
\label{fig1}
\end{figure*}

\begin{figure*}
\includegraphics[width=15cm]{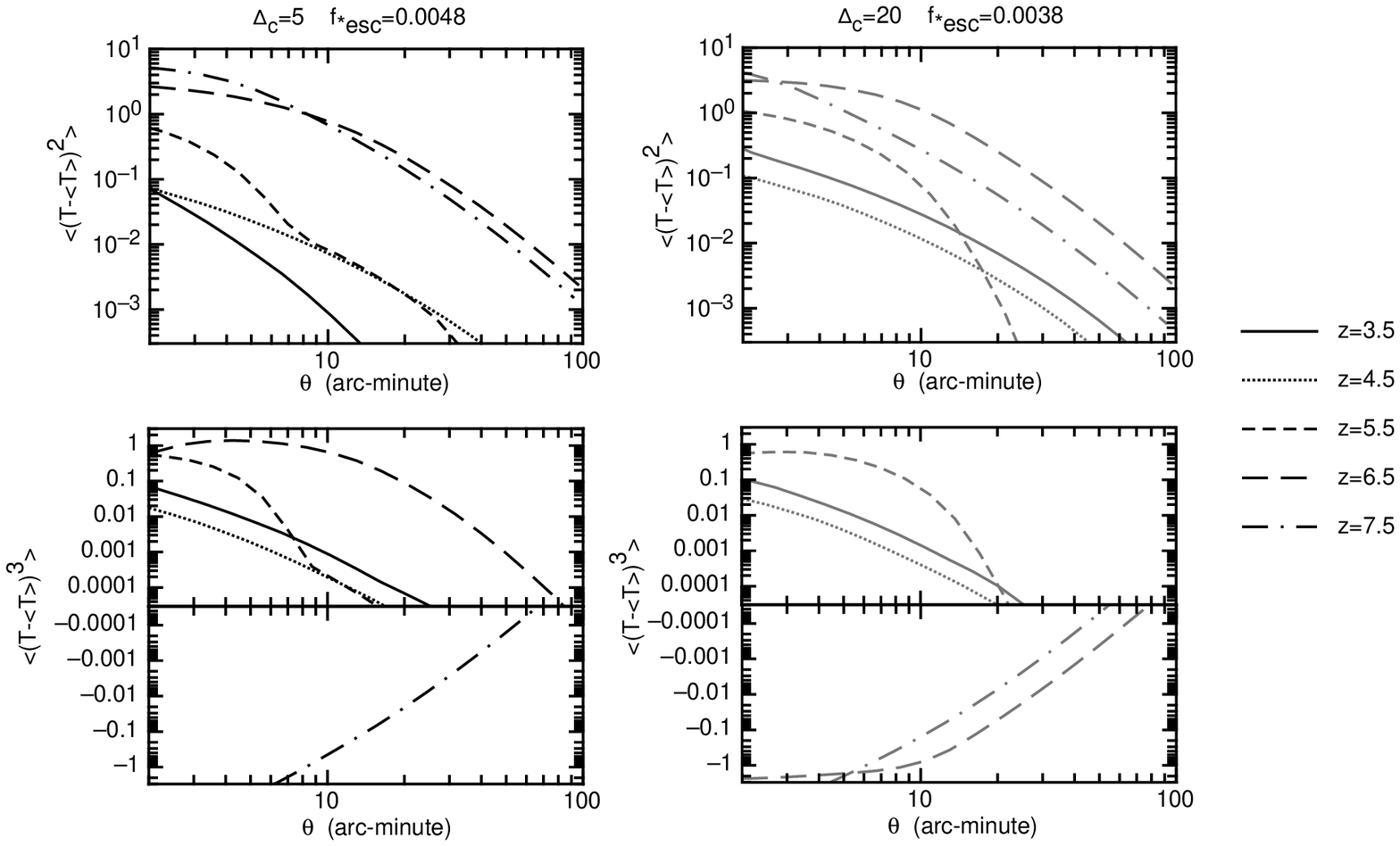} 
\caption{Examples of auto-correlation and skewness functions. \textit{Upper
Panels:} The auto-correlation is plotted as a function of $\theta$ at 5
different redshifts spanning the overlap epoch (for this
model). \textit{Lower Panels:} The corresponding examples of skewness
functions. Two cases are
shown, corresponding to different values for the critical overdensity prior
to the overlap epoch ($\Delta_{\rm c}=5$, left panels; and $\Delta_{\rm c}=20$, right panels).}
\label{fig2}
\end{figure*}

Prior to the redshift of the overlap epoch, the relevant scale $R$ in the
calculation of the biased collapsed fraction (equation~\ref{Fcol}) is equal
to the size of the region being considered. However, following the 
overlap epoch the ionizing photon mean-free-path increases rapidly
and quickly exceeds this scale. Once the mean-free-path is large, the
ionizing photons contributing to the background within a region of size $R$
were not produced locally. At times when the mean-free-path $\lambda$ is
greater than $R$ we therefore assume that a fraction $1-R^3/\lambda^3$ of
the ionizing back-ground within a region of radius $R$ was produced from
sources in the mean-universe (we expect the contribution of the ionizing background from outside $R$  to be only weakly
correlated with the density of the region within $R$ when $\lambda\gg R$). The remaining fraction ($R^3/\lambda^3$) is computed assuming biased
star formation using equation~(\ref{Fcol}). Note that this dilution applies
only to the ionizing background, and not to the density of baryons within
the region of size $R$.

In a cold neutral IGM beyond the redshift of reionization, the collapsed fraction should be computed for halos of sufficient mass to initiate star formation. The critical virial temperature is set by 
the temperature ($T_{{\mathrm{N}}}\sim 10^4$ K) above which efficient atomic hydrogen cooling promotes star formation. Following the reionization of a region, the Jeans mass in the heated IGM limits accretion to halos above $T_{{\mathrm{I}}}\sim10^5$ K (Efstathiou~1992; Thoul \& Weinberg~1996; Dijkstra et al.~2004).
We may therefore write the time derivative of the collapsed fraction 
\begin{align}
\label{collapsed_fraction}
\frac{dF}{ dt_{  {\mathrm{year}}  }}(z)= & 
\left[ Q_{\mathrm{m}}(z)\frac{dF(z,T_{{\mathrm{I}}})}{dz} + [1-Q_{\mathrm{m}}(z)]
\frac{dF(z,T_{\mathrm{N}})}{dz}\right] \notag\\
& \times \frac{dz}{dt_{{\mathrm{year}}}},
\end{align}
where $Q_{\mathrm{m}}$ is the ionized mass fraction in the universe.

To describe the ionizing flux from stars we require one additional
parameter. Only a fraction of ionizing photons produced by stars may enter
the IGM. Therefore an additional factor of $f_{\mathrm{esc}}$ (the escape
fraction) must be included when computing the emissivity of galaxies.
Ciardi \& Ferrara~(2005) include a review of existing constraints on
$f_{\mathrm{esc}}$ which suggests that its value is $\la 15\%$. More recent
simulations (Gnedin, Kravtsov, \& Chen 2007) show the average escape
fraction from galaxies to be only a few percent over a wide redshift
range. The product of the star formation efficiency and the escape fraction
may be combined into a single free parameter ($f_{\rm \star,esc}=f_\star\times f_{\rm esc}$) to
describe the contribution of stars in our model. The ionizing photon
emissivity may then be written as
\begin{equation}
\frac{dn_\gamma}{dt} = f_{\rm \star,esc}N_{\gamma}\frac{dF}{ dt}(z)\rho_{\rm b},
\end{equation} 
where $\rho_{\rm b}$ is the co-moving mass-density of baryons.

\section{Results}

Figure~\ref{fig1} shows example models for the reionization of the IGM and
the subsequent post-overlap evolution of the ionizing radiation field. Two
cases are shown, corresponding to two different values for the critical
overdensity prior to the overlap epoch ($\Delta_{\rm c}=5$ and $\Delta_{\rm
c}=20$), with values of $f_{\rm \star,esc}=0.0048$ and $f_{\rm
\star,esc}=0.0038$ respectively. These cases do not represent the best fit
to the data, but rather bracket the range of the overlap epoch redshifts
($6\la z\la7$) for which our model is consistent with observations at
$z\la6$ (without invoking an additional population of more massive stars at
high redshift; e.g. Wyithe \& Loeb~2003). The examples have different values of
$\Delta_{\rm c}$, and thus illustrate the mild dependence of our results on
this unknown parameter.  We note that the values of $f_{\rm \star,esc}$ required for our
model to reproduce existing observations are in excellent agreement with
external considerations. In particular our value of $f_{\rm \star,esc}\sim$~a~few~$\times10^{-3}$ corresponds to product of recent estimates for the escape
fraction (a few $\times10^{-2}$; Gnedin, Kravtsov, \& Chen 2007), with estimates of the average star-formation rate ($\sim10^{-1}$ from the ratio between the mass density in stars and baryons; Fukugita, Hogan \& Peebles~1998).

In the top left panel of Figure~\ref{fig1} we show the evolution of the ionization rate. The
observational points are from the simulations of Bolton et al.~(2007b; based on the observations of Fan et al.~2006).  In the top-right
panel we plot the corresponding volume (lower curves) and mass (upper
curves) averaged fractions of neutral gas in the universe. Also shown
(dotted lines) is the fraction of the IGM yet to overlap ($1-Q_{\rm i}$). The redshift
where these curves drop to zero is normally quoted as the redshift
of reionization, and these curves correspond approximately to the standard semi-analytic
calculation (e.g. Haiman \& Loeb~1997). However our formalism allows for
the calculation of both mass and volume averaged neutral fractions to lower
redshifts. The observational points for the volume averaged neutral
fraction are from Bolton et al.~(2007b), while the observed mass-fractions
are from the damped Ly$\alpha$ measurements of Prochaska et al.~(2005), and therefore represent lower limits on the total HI content of the IGM. Both
curves show excellent agreement with these quantities, despite their
differing by 3 orders of magnitude. In the lower left panel we plot the
evolution of the ionizing photon mean-free-path. The data points are based
on Storrie-Lombardi et al.~(1994). Again the model is in good agreement
with the available observations. We note that the observed mean-free-path
is found from the number density of Ly-limit systems and is independent of
the Ly$\alpha$ forest absorption derived quantities of ionization rate and volume
averaged neutral fraction, as well as being independent of the HI
mass-density measurements. Our simple model therefore simultaneously reproduces
the evolution of three independent measured quantities.

At a specified redshift, our model yields the mass-averaged fraction
of ionised regions within the IGM on various scales $R$ as a function
of overdensity. We may then calculate the corresponding 21cm
brightness temperature contrast
\begin{equation}
T(\delta,R) = 22\mbox{mK}\left(\frac{1+z}{7.5}\right)^{1/2}(1-Q_{\rm i}F_{\rm M}(\Delta_{\rm i},\delta,R))\left(1+\frac{4}{3}\delta\right),
\end{equation}
where the pre-factor of 4/3 on the overdensity refers to the spherically
averaged enhancement of the brightness temperature due to peculiar
velocities in overdense regions (Bharadwaj \& Ali~2005; Barkana \&
Loeb~2005). In the lower right panel of Figure~\ref{fig1} we show the
evolution of the mean brightness temperature contrast of 21cm emission with
redshift. For comparison, we also show the fraction of IGM yet to
overlap. Significant mean redshifted 21cm emission will extend for at least
a redshift unit beyond overlap (during which time it has dropped by a factor of
$\sim5$). This implies that a global step signature will be very gradual,
making its detection by instruments with a limited band-pass
challenging.

\subsection{Fluctuations in 21cm emission}

Given the distribution of $\delta$ from the primordial power-spectrum of
density fluctuations, we may find the probability distribution $dP/dT$ of
brightness temperature $T$ in redshifted 21cm intensity maps, and hence the
second and third moments of the distribution as functions of angular scale
and redshift (Wyithe \& Morales~2007).  The structure of our semi-analytic
model for reionisation makes it natural to discuss the moments of the real
space intensity fluctuations smoothed within top-hat window functions of
angular radius $\theta$.

The second moment of these distributions corresponds to the auto-correlation function of brightness temperature
smoothed on an angular radius $\theta$:
\begin{eqnarray}
\nonumber
\langle \left(T-\langle T\rangle\right)^{2}\rangle &=&\\
 &&\hspace{-15mm}\left[\frac{1}{\sqrt{2\pi}\sigma(R)}\int d\delta \left(T(\delta,R)-\langle T\rangle\right)^2e^{-\frac{\delta^2}{2\sigma(R)^2}} \right],
\end{eqnarray}
where 
\begin{equation}
\langle T\rangle = \frac{1}{\sqrt{2\pi}\sigma(R)}\int d\delta~ T(\delta,R)e^{-\frac{\delta^2}{2\sigma(R)^2}} ,
\end{equation}
and $\sigma(R)$ is the variance of the density field (at redshift $z$) smoothed on a scale
$R$.

\begin{figure*}
\includegraphics[width=15cm]{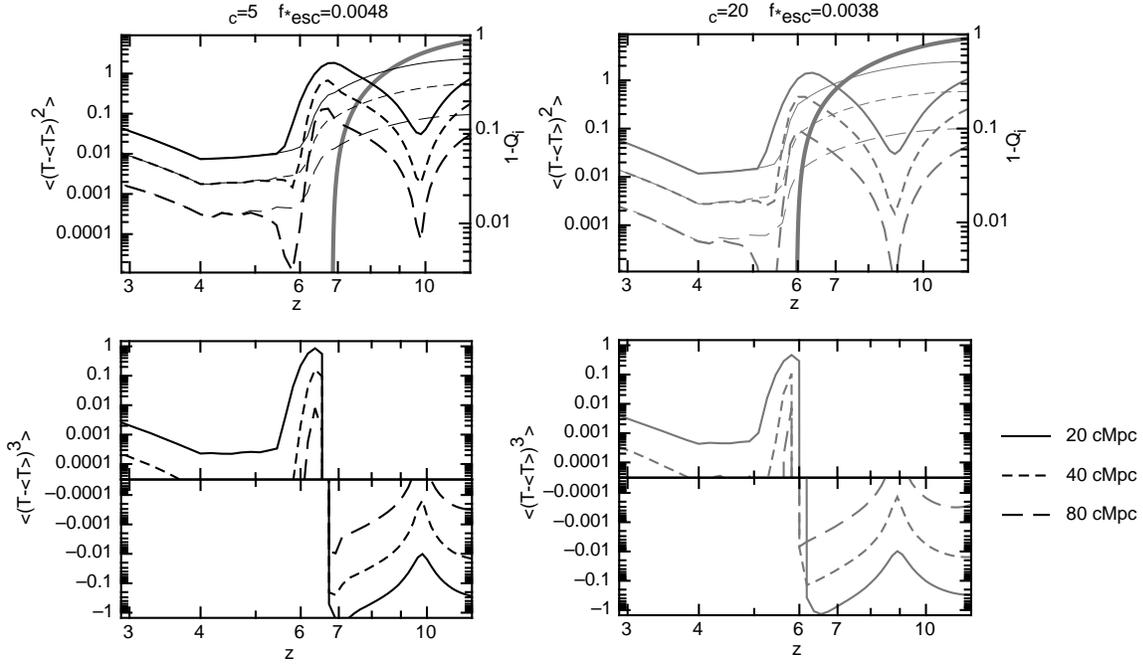} 
\caption{Examples of auto-correlation and skewness functions.  Two cases are
shown, corresponding to different values for the critical overdensity prior
to the overlap epoch ($\Delta_{\rm c}=5$, left panels; and $\Delta_{\rm c}=20$, right panels). \textit{Upper
Panels:} The auto-correlation is plotted as a function of redshift for 3
different co-moving scales of relevance for observations using upcoming low-frequency radio telescopes. The heavy lines show our
model. For comparison the thin lines show the fluctuations in the absence
of galaxy bias. Also shown for comparison is the fraction of IGM yet to
overlap (thick grey line dropping to zero at $z=6-7$). \textit{Lower Panels:} The
corresponding examples of skewness functions.}
\label{fig3}
\end{figure*}

A measure of the departure of the statistics from those of a Gaussian
random field is provided by the third moment of the distribution
$dP/dT$. We refer to the third moment as the skewness function
 of brightness temperature smoothed with top-hat
windows of angular radius $\theta$:
\begin{eqnarray}
\nonumber
\langle \left(T-\langle T\rangle\right)^{3}\rangle&=&\\
 &&\hspace{-15mm} \left[\frac{1}{\sqrt{2\pi}\sigma(R)}\int d\delta
 \left(T(\delta,R)-\langle
 T\rangle\right)^3e^{-\frac{\delta^2}{2\sigma(R)^2}}
 \right].
\end{eqnarray}

Examples of auto-correlation and skewness functions are plotted in
Figure~\ref{fig2}. The upper panels show the auto-correlation plotted as a
function of $\theta$ at 5 different redshifts spanning the
overlap epoch (for this model). The lower panels show the corresponding
examples of skewness functions. Further examples of auto-correlation and
skewness functions are plotted in Figure~\ref{fig3}, but this time as a
function of redshift for 3 different co-moving scales of relevance for upcoming low-frequency radio telescopes. 
The thick dark lines show the evolution for our full model. We also show
(thin dark lines) the fluctuations that would be present in the absence of galaxy bias. Comparison of these curves demonstrates that galaxy
bias is important in setting fluctuation levels prior to the overlap epoch,
but becomes progressively less important after the overlap epoch is completed. 
This is because the
mean-free-path becomes large following the completion of overlap, and as a result the ionizing photons that form
the background at a particular location in the IGM are then drawn from a volume that is much larger
than the regions among which the fluctuations are measured. Galaxy bias remains important until lower redshifts on larger scales $R$ because the mean-free-path exceeds $R$ only at later times. Also shown for comparison (heavy grey lines) is the
fraction of IGM yet to overlap ($1-Q_{\rm i}$).

Figures~\ref{fig2}-\ref{fig3} suggest that a characteristic feature of the overlap epoch will be the
inversion of the 21cm intensity distribution from negative to positive skewness. As mentioned above, the ionization state of an
overdense region is set by the competing effects of galaxy bias, and higher
gas density (and recombination rate). With respect to 21cm fluctuations, this inversion therefore corresponds
to the shift in dominance from the biased radiation field prior to the
overlap epoch, to dominance of gas density and recombinations on small
scales after the overlap epoch (when the long ionizing mean-free path
washes out the effect of galaxy bias).

\begin{figure*}
\includegraphics[width=15cm]{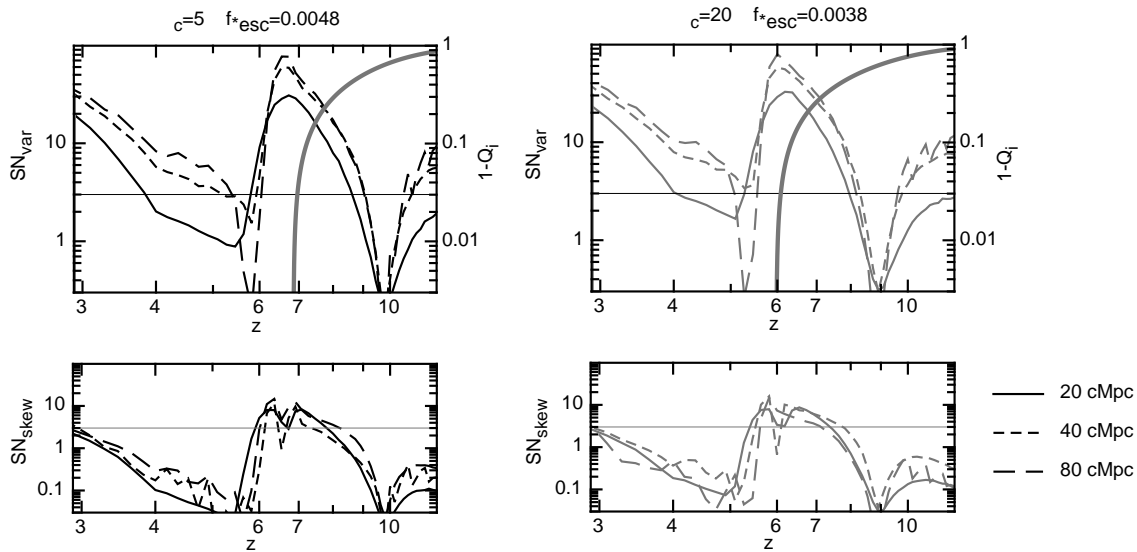} 
\caption{Examples of signal-to-noise for auto-correlation and skewness
functions measured with the MWA assuming 1000 hours integration.  Two cases are
shown, corresponding to different values for the critical overdensity prior
to the overlap epoch ($\Delta_{\rm c}=5$, left panels; and $\Delta_{\rm c}=20$, right panels). \textit{Upper
Panels:} The signal-to-noise for the auto-correlation function is plotted
as a function of redshift for 3 different co-moving scales of relevance for
observations using upcoming low-frequency radio telescopes. Shown for comparison is the fraction of IGM yet to overlap according
to the definition in \S~\ref{models} (thick grey lines). \textit{Lower Panels:} The corresponding examples of
signal-to-noise for skewness functions. Horizontal lines have been drawn at
a signal-to-noise of 3 to guide the eye.}
\label{fig4}
\end{figure*}

To investigate the sensitivity of forthcoming low-frequency telescopes to
the fluctuations in 21cm emission after reionization, we have estimated the
signal-to-noise for the Mileura-Widefield Array (MWA\footnote{see
http://www.haystack.mit.edu/ast/arrays/mwa/index.html}) using the formalism
outlined in Wyithe \& Morales~(2007). The MWA, which is currently under
construction will comprise a phased array of 500 tiles (each tile will
contain 16 cross-dipoles) distributed over an area with diameter 1.5km.  In
Figure~\ref{fig4} we show examples of signal-to-noise for auto-correlation
and skewness functions measured with the MWA assuming 1000 hours
integration (the examples correspond to the models in
Figure~\ref{fig3}). In the upper panels we plot the signal-to-noise for the
auto-correlation as a function of redshift for 3 different co-moving scales
of relevance for the MWA. Also shown for comparison as before is the
fraction of IGM yet to overlap ($1-Q_{\rm i}$). In the lower panels we show
the corresponding signal-to-noise for the skewness function. The model
suggests that the auto-correlation will reach a minimum around a redshift
unit following the completion of the overlap epoch, but that even here it
will remain detectable at a signal-to-noise of several. Moreover the
signal-to-noise at $z\sim3-4$ should be comparable with that near the peak
of the reionization era due to the decreased level of sky-noise at lower
redshifts. The MWA is designed to be sensitive to frequencies as high as
300MHz, or redshifts as low as $z\sim3.5$. Thus redshifted 21cm
observations could be used to track the entire overlap era into the post
reionization IGM.

\begin{figure*}
\includegraphics[width=17cm]{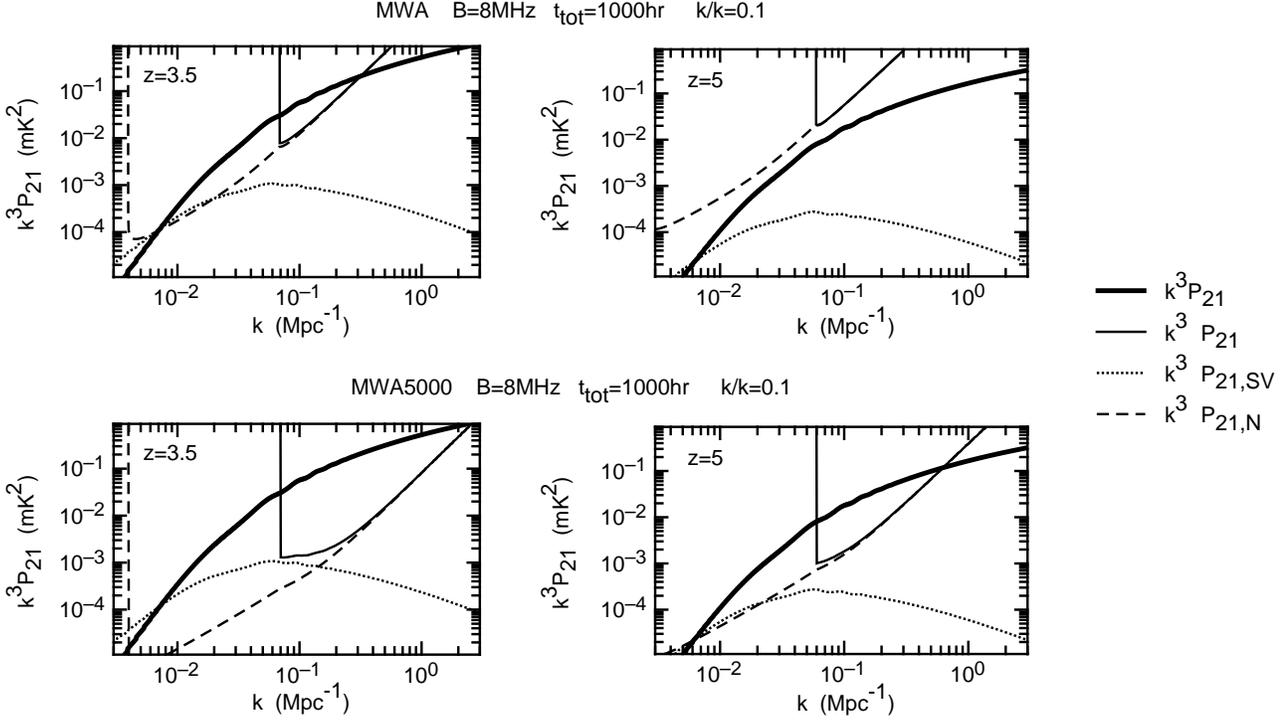} 
\caption{The power-spectrum of 21cm fluctuations after reionization (thick
solid lines) at $z=3.5$ (left panels) and $z=5$ (right panels). The model
power-spectrum shown was computed assuming $\Delta_{\rm c}=20$.  Also shown
for comparison are estimates of the noise for MWA (upper panels) and its
future extension MWA5000 (lower panels). In each case we plot the
sample-variance (dotted lines) and thermal noise (dashed lines) components
of the uncertainty within $k$-space bins of size $\Delta k=k/10$. The
combined uncertainty including the minimum $k$ cutoff due to foreground
subtraction, is also shown as the thin solid lines.}
\label{fig5}
\end{figure*}

\section{The power-spectrum of 21cm fluctuations after reionization}

Our analysis thus far has concentrated on the auto-correlation function of
21cm fluctuations. A more powerful statistical probe will be provided by
the power-spectrum which is naturally accessible to interferometric
observations such as those to be carried out by the MWA. During the
reionization epoch, the relation between the power-spectrum of 21cm
fluctuations and the underlying matter power-spectrum is complex, and in
the late stages is dominated by the formation of large ionized bubbles
(Furlanetto et al.~2004; McQuinn et al.~2006). After reionization
(i.e. $z\la5.5$) there are no longer separate ionized bubbles as most of
the IGM is ionized. Our modeling suggests that the skewness of the 21cm
intensity fluctuation distribution will be small during the post overlap
epoch. As a result the 21cm emission will follow the density field up to a
constant of proportionality whose value reflects the neutral fraction and a
bias like factor due to the varying rate of recombination with
overdensity. Thus, after the overlap of HII regions we may write the
following expression for the power-spectrum of 21cm fluctuations
\begin{equation}
P_{21}(k)=b_{21}^2P(k)D(z)^2,
\end{equation}
where $P(k)$ is the primordial power-spectrum of the density field,
extrapolated linearly to $z=0$. The bias $b_{21}$, which has dimensions of temperature squared
may be determined directly from our calculation of the auto-correlation
function, i.e.
\begin{equation}
b_{21}(R)^2=\frac{\langle(T-\langle T\rangle)^2\rangle}{\sigma(R)^2}.
\end{equation} 
The bias $b_{21}$ is a function of $R$, but at $z\la5$ is nearly
independent of scale over the range of interest. We show examples of the
resulting power-spectrum (thick solid lines) in Figure~\ref{fig5} at
$z=3.5$ (left panels) and $z=5$ (right panels). The model power-spectrum
shown was computed assuming $\Delta_{\rm c}=20$, for which we find
$b_{21}\sim1.2$mK at both $z=3.5$ and $z=5$.

Calculations of the sensitivity to the 21cm power-spectrum for an
interferometer have been presented by a number of authors. We follow the
procedure outlined by McQuinn et al.~(2006), drawing on results from
Bowman, Morales \& Hewitt~(2006) for the dependence of the array antenna
density on radius $\rho(r)$.  The uncertainty in
a measurement of the power-spectrum has two separate components, due to the
thermal noise of the instrument ($\Delta P_{\rm 21,N}$), and due to sample
variance within the finite volume of the survey ($\Delta P_{\rm
21,SV}$). The contamination of foregrounds provides an additional source of
uncertainty in the estimate of the power-spectrum. McQuinn et al.~(2006)
have shown that it should be possible to remove the power due to
foregrounds to a level below the cosmological signal, provided that the
region of band-pass from which the power-spectrum is estimated is
substantially smaller than the total band-pass available. Following the
approximation suggested in McQuinn et al.~(2006), we combine the above
components to yield the uncertainty on the estimate of the power-spectrum
\begin{eqnarray}
\nonumber
\Delta P_{21} &=& \Delta P_{\rm 21,SV} + \Delta P_{\rm 21,N}\hspace{3mm}\mbox{if $k>k_{\rm min}$}\\
              &=& \infty \hspace{26mm}\mbox{otherwise}
\end{eqnarray}
where $k_{\rm min}=2\pi/y$ and $y$ is the co-moving line-of-sight distance corresponding to the band-pass within which the power-spectrum is measured.

Estimates of the sample-variance (dotted lines) and thermal noise (dashed
lines) components of the uncertainty for detection by the MWA are plotted
in the upper panels of Figure~\ref{fig5}. We model the antennae distribution as having $\rho(r)\propto r^{-2}$ with a maximum radius of 750m and a finite density core of radius 20m, and we assume a 1000hr integration, a bandpass of $B=8$MHz,
and $k$-space bins of width\footnote{The signal-to-noise is increased in proportion to $\sqrt{\Delta k}$, and so will be substantially better per bin in measurements of the power-spectrum at lower resolution in $k$.} $\Delta k=k/10$. The combined uncertainty
including the minimum $k$ cutoff due to foreground subtraction is shown as
the thin solid line. The results confirm our expectations based on results
in the previous section. The power-spectrum of 21cm fluctuation should
be detectable at high significance at $z\sim3.5$, well after the completion
of reionization. Indeed the signal-to-noise for detection of the
power-spectrum at $z\sim3.5$ is not dissimilar to expectations for the
signal-to-noise during the reionization era itself (e.g. McQuinn et
al.~2006). On the other hand, due to the increasing sky dominated
system-noise at higher redshift, a detection will be more challenging at
redshifts just after overlap. 

At values of $k\sim$ a few $\times10^{-1}$ Mpc$^{-1}$, the measurement of
the power-spectrum using the MWA will be limited by the thermal sensitivity
of the array, and so the signal-to-noise achievable in this regime will be
greatly enhanced by a subsequent generation of instruments with larger
collecting area. As an example we consider a hypothetical followup
instrument to the MWA which would comprise 10 times the total collecting
area.  We refer to this followup telescope as the MWA5000. The design
philosophy for the MWA5000 would be similar to the MWA, and we therefore
assume antennae distributed as $\rho(r)\propto r^{-2}$ with a diameter of
2km and a flat density core of radius 80m (see McQuinn et al.~2006). In the
lower panels of Figure~\ref{fig5} we present estimates for measurement of
the 21cm power-spectrum using MWA5000. The sample-variance (dotted
lines) and thermal noise (dashed lines) components of the uncertainty are
plotted as before assuming a 1000hr integration, a bandpass of $B=8$MHz,
and $k$-space bins of width $\Delta k=k/10$. The combined uncertainty
including the minimum $k$ cutoff due to foreground subtraction is shown as
the thin solid line. This figure, in combination with previous calculations
of sensitivity to the 21cm power-spectrum during reionization (e.g. McQuinn
et al.~2006) demonstrates that the MWA5000 could measure the 21cm
power-spectrum in detail across the entire frequency range accessible to an
MWA antennae, corresponding to a redshift range of $3.5\la z\la15$.

The reionization epoch has formed the dominant driver for studies of the
statistical properties of redshifted 21cm emission to date, while we have argued in this paper that the study of the IGM using 21cm emission could also be extended to lower post-reionization redshifts. However the measurement
of a 21cm power spectrum will also provide significant cosmological
information (McQuinn et al.~2006). For example the angular scale of the
baryonic acoustic peak in the matter power spectrum, measured through
redshifted 21cm observations could be used to constrain the evolution of
the dark energy with cosmic time (Blake \& Glazebrook~2003; Seo \&
Eisenstein 2007; Angulo et al. 2007) in the redshift range of $3.5\la z \la
6$, as well as during the reionization era.  No other probes are currently
effective during these epochs (Corasaniti, Huterer, \& Melchiorri 2007). A
detailed analysis of the prospects for such a study with future
low-frequency arrays will be presented elsewhere.

\section{Discussion}

While the recent discovery of distant quasars has allowed detailed
absorption studies of the state of the IGM at $z\sim 6$ (Fan et al.~2006;
White et al.~2003), the interpretation of the data is not yet robust. The
prime reason for the ambiguity in interpreting high redshift quasar
absorption spectra is that Ly$\alpha$ absorption can only be used to probe
volume averaged neutral fractions that are smaller than $10^{-3}$, owing to
the large cross-section of the Ly$\alpha$ resonance. A better probe of the
process of reionization is provided by redshifted 21cm observations. The
conventional wisdom is that the 21cm signal disappears after the {\it
overlap} epoch, because there is little neutral hydrogen left through most
of the intergalactic space. But observations of damped Ly$\alpha$ systems
out to redshift $z\sim4$ show the density parameter of HI to be
$\Omega_{\rm HI}\sim10^{-3}$, indicating that the mass averaged neutral
hydrogen fraction remains at the level of a few percent throughout cosmic
history.

Existing constraints from Ly$\alpha$ studies are based on line-of-sight
averaged optical depths and are therefore sensitive to the volume filling
fraction of neutral hydrogen. These observations are limited in their
ability to probe the IGM once the overlap era is approached due to
the high optical depth to Ly$\alpha$ absorption.  On the other hand,
redshifted 21cm observations directly probe the pockets where
most of the hydrogen resides after reionization.
In this paper we have considered the detection of fluctuations in 21cm
emission from neutral hydrogen after the completion of reionization using
low-frequency telescopes that are currently under construction (specifically the Mileura Widefield Array, MWA).  Based on a
physically motivated model for the reionization era we have shown that the
residual neutral gas, which is contained in high density pockets following the end of the
overlap epoch, produces a substantial signal of 21cm fluctuations. These
fluctuations directly probe the post-reionization IGM and have a power-spectrum that will be detectable by the MWA at a
signal-to-noise ratio that is comparable to observations during the
reionization epoch.  We stress that in the case of observations at $z\la5$ the
prediction of the power-spectrum amplitude is quite robust, being primarily
dependent on the underlying mass power-spectrum and the mass-averaged neutral
fraction of hydrogen which, unlike predictions for the reionization epoch,
represent well-measured quantities. 

Models of reionization predict that non-Gaussianity in the distribution of
21cm intensities would be the signature of star-formation driven reionization
of the IGM. Our calculations predict that after the overlap of HII
regions in the IGM, there will be a transition between an epoch of negative
skewness when the statistics of the intensity distribution are driven by
galaxy-bias (prior to overlap), and an epoch of positive skewness when
they are driven by the density distribution and recombination rate
(post-overlap). Observation of this rapid transition in the statistics of
the 21cm signal will therefore provide added confidence that the end of the
reionization era has been detected.

Observations of the post-reionization 21cm power-spectrum will be significantly enhanced by 2nd generation low-frequency telescopes with increased collecting area. In addition to the utility of 21cm fluctuations for studies of the evolution in the ionization state of the IGM during the post reionization epoch, measurement of the 21cm power-spectrum may therefore yield precise measurements of the angular scale of
the acoustic peak in the matter power-spectrum. Thus in the future it may become possible to use redshifted 21cm studies to constrain the
dark energy in the unexplored redshift range of $3.5\la z\la 15$, where it is not accessible through other techniques.

\bigskip

{\bf Acknowledgments} The research was supported by the Australian Research
Council (JSBW) and Harvard University grants (AL). AL thanks 
the astrophysicists at Caltech for their kind hospitality
as the Kingsley visitor during the completion of this work.

\newcommand{\noopsort}[1]{}

\label{lastpage}
\end{document}